# Rapid and Scalable Synthesis of Alkali Metal-Intercalated $C_{60}$ Superconductors

A Iyo,[1*] H Fujihisa,[1] Y Gotoh,[1] S Ishida,[1] H Eisaki[1], H Ogino[1], K Kawashima[1,2]

[1]National Institute of Advanced Industrial Science and Technology (AIST), Tsukuba, Ibaraki 305-8568, Japan

[2] IMRA JAPAN Co., Ltd., Kariya, Aichi 448-8650, Japan

*E-mail: iyo-akira@aist.go.jp

**Abstract**

Alkali metal-intercalated $C_{60}$, $A_3C_{60}$ ($A$ = K, Rb, Cs, and their combinations), holds significant potential for practical applications due to its high superconducting transition temperature (33 K), high upper critical field (900 kOe), and isotropic superconductivity. However, application-oriented research has been limited by the lack of an efficient $A_3C_{60}$ synthesis process. In this study, we demonstrate a rapid and scalable synthesis of $A_3C_{60}$ ($A$ = K, Rb, and $Cs_{1/3}Rb_{2/3}$) via direct mixing of $A$ and $C_{60}$, realizing the fabrication of high-quality sintered $A_3C_{60}$ pellets within just 1 hour of heating at 200–300°C. The pellets exhibited large superconducting shielding volume fractions with sharp transitions, and the relationship between the lattice constant and transition temperature was in good agreement with previous reports. This direct mixing method enables simple and rapid production of large quantities of $A_3C_{60}$, which is expected to accelerate research into applications such as superconducting wires and bulk magnets.

Keywords: Alkali metal-intercalated $C_{60}$, $A_3C_{60}$ ($A$ = K, Rb, $Cs_{1/3}Rb_{2/3}$), $A_6C_{60}$ ($A$ = K, Rb, Cs), superconductor, rapid and scalable synthesis, direct mixing method

**1. Introduction**

Following the discovery of superconductivity at 18 K in potassium (K)-intercalated fullerene ($C_{60}$), $K_3C_{60}$ [1], extensive research efforts have been directed toward the exploration of alkali-metal ($A$)-intercalated $C_{60}$, leading to the synthesis of new superconducting compounds such as $A_3C_{60}$ ($A$ = Rb, Cs, and their combinations) [2–4]. Among these, rubidium (Rb)- and cesium (Cs)-intercalated variants exhibit an increase in the superconducting transition temperature ($T_c$), reaching a maximum of 33 K [5-7]. This trend, which correlates with the expansion of the cubic unit cell upon intercalation with larger alkali metals, highlights the crucial role of lattice structure and electronic band modifications in tuning the superconducting properties of $A_3C_{60}$ [7–9].

One of the most remarkable characteristics of $A_3C_{60}$ superconductors is their exceptionally high upper critical field ($H_{c2}$), reaching as much as 900 kOe [10–12], which is among the highest values reported for cubic-structured superconductors. This property suggests the potential for robust superconductivity even under strong external magnetic fields, making these materials attractive candidates for high-field applications. Unlike high-temperature cuprate or iron-based superconductors, which exhibit significant anisotropy due to their layered structures [13,14], the isotropic nature of superconductivity in $A_3C_{60}$ arises from conventional phonon-mediated pairing mechanisms [15,16]. This aspect is particularly advantageous for technological applications such as superconducting wires and bulk magnets, as it eliminates the need for precise crystal orientation control. Despite its high potential, research and development of $A_3C_{60}$ have been limited due to the lack of an efficient production process.

$A$-intercalated $C_{60}$ exists in several distinct phases, $A_xC_{60}$ ($x$ = 3, 4, and 6) [17]. The physical properties of $A_xC_{60}$ are highly sensitive to the amount of intercalation ($x$) [18–20], and superconductivity emerges exclusively in $A_3C_{60}$. Therefore, precise control of $x$ is critical to obtain materials with excellent superconducting properties. The vapor-solid phase reaction method is the most established [21,22], involving $A$ vapor, generated within a sealed container by heating, reacting with solid-phase $C_{60}$ in a sealed container. The reaction progresses slowly from the surface of $C_{60}$ in contact with $A$ vapor, necessitating prolonged heat treatment (typically several days). Solution-phase methods using ammonia or alkylamine with dissolved alkali metals [23–27] have also been developed but remain impractical due to slow reaction rates, difficulty in achieving the required stoichiometry, and the need for post-treatment to remove solvents. Other approaches, such as electrochemical methods and self-propagating high temperature synthesis [28–32], are also face limitations for extension to a practical rapid mass production process of $A_3C_{60}$.

In this study, we propose a practical method for synthesizing $A_3C_{60}$ using a precursor obtained by directly mixing $A$ and $C_{60}$. Since $A$ is already intercalated into $C_{60}$, albeit heterogeneously, through the mixing process, $A_3C_{60}$ forms within just 1 hour of heating at 200–300°C. This rapid formation eliminates the need for sealed containers to confine the $A$ vapor for an extended period, unlike the conventional vapor-solid-phase reaction method. This method also allows precise control of the amount of $A$ intercalated into $C_{60}$ through simple weighing and enables the production of sintered samples that are challenging to obtain with conventional methods. Thus, the direct mixing method offers a simple, rapid, and scalable alternative to conventional methods.

## 2. Experimental

*2.1 Synthesis process of $A_3C_{60}$ and $A_6C_{60}$*

$A_3C_{60}$ polycrystalline samples were synthesized following the process outlined in **figures 1(**a**)**–(d), using K (99.5%, Sigma-Aldrich), Rb (99.9%, Kojundo Chemical Lab.), Cs (99.98%, Furuuchi Chemical), and $C_{60}$ powder (99.5%, BBS Chemicals) as starting materials. In this study, we focused on $A_3C_{60}$ ($A$ = K, Rb, and $Cs_{1/3}Rb_{2/3}$), which are representative compositions among intercalated $C_{60}$ superconductors that have been studied extensively due to their high $T_c$. Note that $Cs_3C_{60}$ does not exhibit superconductivity at ambient pressure, but superconductivity at 38 K is induced by applying pressure [33]. The process was conducted in an argon-filled glove box. $A$ and $C_{60}$ were weighed in a 3:1 molar ratio, totaling approximately 0.2 g, and mixed for about 15 min using a zirconia mortar and pestle (**figures 1(**a**)** and **1(**b**)**).

The mixture was pressed into pellets and heated at $T_h$ = 100–500°C for 1 h using an electric furnace inside the glove box. Due to the short heating duration, without the use of sealed containers to confine the $A$ vapor, negligible mass loss was observed even at 500°C. To minimize potential reactions with trace oxygen and moisture present in the glove box and prevent contamination, the sample was wrapped in aluminum foil during heating (**figure 1(**c**)**). **Figure 1**(d) shows the sintered $K_3C_{60}$ pellet that had densities of approximately 70–80% of the theoretical value.

For comparison with the synthesis of $A_3C_{60}$, we also attempted to synthesize the alkali metal-saturated phase $A_6C_{60}$ ($A$ = K, Rb, and Cs). $A_6C_{60}$ samples were prepared by simply mixing $A$ and $C_{60}$ in a 6:1 molar ratio. **Figures 1**(e) and **1**(f) depict crystal structures of $A_3C_{60}$ and $A_6C_{60}$, respectively.

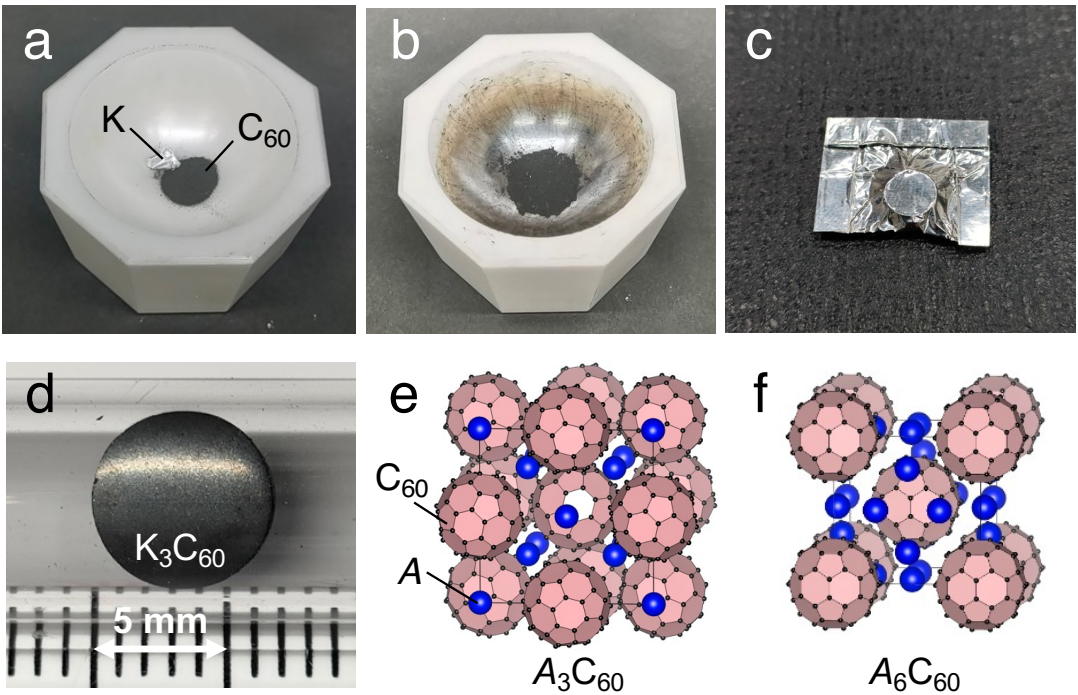

**Figure 1**. Schematic overview of the $A_3C_{60}$ synthesis process used in this study, exemplified by $K_3C_{60}$. (a)–(b) K and $C_{60}$ were weighed (totaling approximately 0.2 g) and mixed with a mortar for around 15 min. (c) The mixture was pressed into a pellet, wrapped in aluminum foil, and heated in an electric furnace inside the glove box. (d) $K_3C_{60}$ sintered pellet obtained after heating at 500°C for 1 h. (e)–(f) Crystal structures of $A_3C_{60}$ and $A_6C_{60}$, respectively.

*2.2. Measurements*

Samples were evaluated by powder X-ray diffraction (XRD) with CuKα radiation (Rigaku Ultima IV). To ensure comparability of diffraction peak intensities, the X-ray irradiation area was standardized across samples. XRD measurements were performed in an airtight setup to avoid exposure, as $A_xC_{60}$ is highly reactive to oxygen and moisture. The lattice constants were refined by the Pawley refinement [34] using Materials Studio Reflex (version 2024 SP1 HF2) [35]. Temperature ($T$) dependence of magnetization ($M$) was measured in zero-field-cooled (ZFC) and field-cooled (FC) modes in a magnetic field ($H$) of 10 Oe using a magnetic property measurement system (Quantum Design, MPMS-XL7). The magnetic field was applied from the transverse direction of the cylindrical pellet-shaped sample. The measured magnetization was corrected using the demagnetization factor ($N$). For example, for a typical pellet size of 4.3 mm in diameter and 1.1 mm in thickness prepared in this study, $N$ was calculated to be 0.21 [36]. The $T$-dependence of electrical resistivity ($\rho$) was measured by a four-terminal method using a physical property measurement system (Quantum Design, PPMS). Electrodes were attached to the sample using silver paste (Dupont 4922N) in a glove box and the sample was covered with grease (Apiezon N) to avoid contact with air.

## 3. Results and discussion

*3.1. Formation of $A_3C_{60}$*

**Figures 2**(a)–(c) display the powder XRD patterns of samples with nominal compositions $A_3C_{60}$ ($A$ = K, Rb, and $Cs_{1/3}Rb_{2/3}$), respectively, as a function of $T_h$ (= 100, 200, 300, 400, and 500°C). For $A$ = K and Rb, $A_3C_{60}$ had already formed as the main phase in the as-mixed samples, with no diffraction peaks corresponding to unreacted $A$ and $C_{60}$, indicating that $A$ and $C_{60}$ almost reacted during mixing. In contrast, for $A$ = $Cs_{1/3}Rb_{2/3}$, the as-mixed sample showed smaller diffraction peaks primarily from $(Cs,Rb)_6C_{60}$ and $C_{60}$, with minimal $CsRb_2C_{60}$ formation even at $T_h$ = 100°C. The temperature required for $CsRb_2C_{60}$ formation was higher than for $A_3C_{60}$ ($A$ = K and Rb), likely due to the two types of $A$ atoms occupying a single atomic site in $CsRb_2C_{60}$, hindering diffusion of $A$ for homogeneous crystallization. The diffraction peak intensity of $A_3C_{60}$ remained largely unchanged from $T_h$ = 200°C (300°C) up to $T_h$ = 500°C for $A$ = K and Rb ($Cs_{1/3}Rb_{2/3}$), indicating that the reaction completed after 1 h heat treatment at 200°C (300°C). Since $A$ was already intercalated into $C_{60}$, albeit non-uniformly, even minimal diffusion of $A$ due to the short heating duration may have resulted in the formation of $A_3C_{60}$. The lattice constants of $A_3C_{60}$ ($A$ = K, Rb, and $Cs_{1/3}Rb_{2/3}$) were determined to be 14.253(1) Å, 14.426(1) Å, and 14.458(1) Å, respectively.

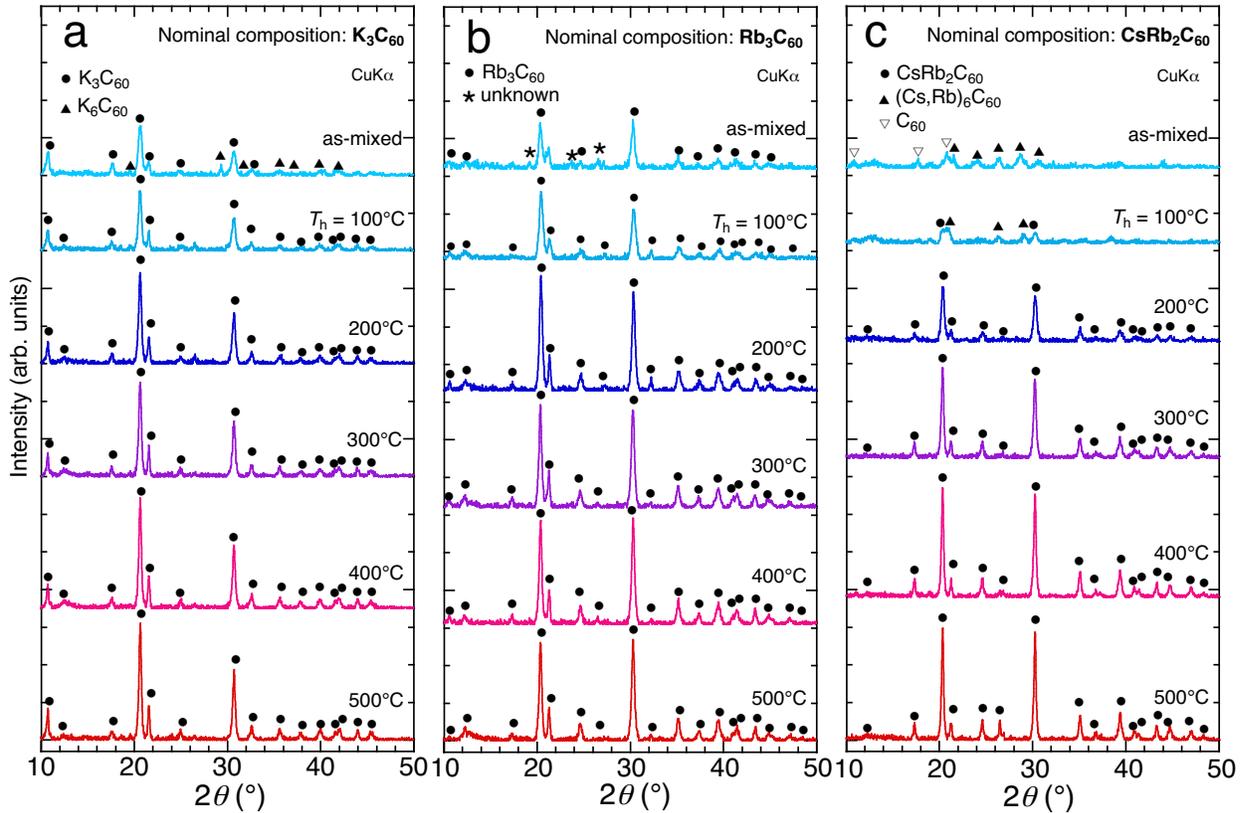

**Figure 2.** (a)–(c) Powder XRD patterns of samples with nominal compositions $A_3C_{60}$ ($A$ = K, Rb, and $Cs_{1/3}Rb_{2/3}$), respectively. The samples were heated at $T_h$ (= 100, 200, 300, 400, and 500°C) for 1 h. Peaks corresponding to $A_3C_{60}$ are marked with black circle.

### 3.2. Formation of $A_6C_{60}$

The powder XRD patterns of as-mixed and heated (at 300°C for 1 h) samples with nominal compositions of $A$:$C_{60}$ = 6:1 ($A$ = K, Rb, and Cs) are shown in **figure 3**. All diffraction peaks were assignable to $A_6C_{60}$. The diffraction peak intensity did not significantly increase after heat treatment, indicating that the reaction of $A$ and $C_{60}$ was complete by mixing alone. This is because $A_6C_{60}$, whose crystal structure is shown in **figure 1(**f**)**, is an alkali metal saturated phase and, unlike $A_3C_{60}$, does not require $A$ diffusion for crystallization by heat treatment. To the best of our knowledge, this is the first study to show that $A_6C_{60}$ can be synthesized without heating. The lattice constants of $A_6C_{60}$ ($A$ = K, Rb, and Cs) were determined to be 11.380(1) Å, 11.546(1) Å, and 11.793(1) Å, respectively, in excellent agreement with reported values [17].

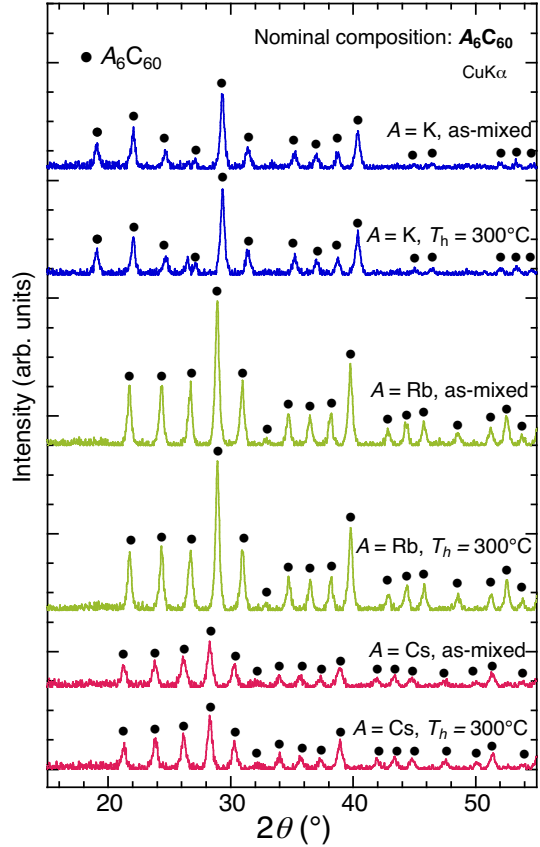

**Figure 3.** Powder XRD patterns of as-mixed and heated ($T_h$ = 300°C for 1 h) samples with nominal compositions of $A$:$C_{60}$ = 6:1 ($A$ = K, Rb, and Cs). Peaks corresponding to $A_6C_{60}$ are marked with black circles.

*3.3 Evaluation of superconductivity in $A_3C_{60}$*

**Figures 4**(a)–(c) show the temperature dependence of $4\pi M/H$ for $K_3C_{60}$, $Rb_3C_{60}$, and $CsRb_2C_{60}$ pellets heated at 500°C for 1 h. All samples exhibited diamagnetic transitions due to superconductivity, with shielding volume fractions of approximately 90% ($4\pi M/H \simeq -0.9$ at low temperatures in the ZFC curves), among the highest reported. The insets display sharp transitions at 19.0 K, 29.6 K, and 32.0 K for $K_3C_{60}$, $Rb_3C_{60}$, and $CsRb_2C_{60}$, respectively, in the FC magnetization curves. These large volume fractions and sharp transitions indicate that the samples are well-sintered and homogeneous. The step-like behavior observed just below $T_c$ in the ZFC magnetization curves of all samples likely results from intergranular weak links observed in superconducting polycrystals [37].

The $\rho$ of $A_3C_{60}$, typically measured in single crystals or thin films [10, 38–40], can be easily assessed using pellet samples. **Figure 4**(d) shows the temperature dependence of $\rho$ near the superconducting transitions measured for $K_3C_{60}$, $Rb_3C_{60}$, and $CsRb_2C_{60}$ pellets. As indicated in figure 4(d), $T_c^{onset}$ and $T_c^{end}$ are defined as the temperatures at which the linear extrapolation of the transition intersects the baseline. Table 1 summarizes $T_c^{onset}$, $\Delta T_c$ (= $T_c^{onset}$ - $T_c^{end}$) and the lattice constants of $A_3C_{60}$ and $A_3C_{60}$ synthesized in this study. Sharp superconducting transitions with a width of approximately 2 K were observed at temperatures nearly identical to those obtained from magnetization measurements.

The sinterability of $A_3C_{60}$ obtained by this method is particularly advantageous for fabricating superconducting wires and bulk magnets [12]. For example, wire fabrication by the powder-in-tube method would benefit from this feature, as the as-mixed powder within the wire can be sintered with a short heat treatment up to 500°C, facilitating the production of superconducting coils through the wind-and-react process [41].

**Figure 4(**e) summarizes the relationship between the lattice constant and $T_c$ for $K_3C_{60}$, $Rb_3C_{60}$, and $CsRb_2C_{60}$ obtained in this study, along with reported data [4,21,22,26]. The good agreement between the present and previously studies confirm that the amount of $A$ intercalation into $C_{60}$ is accurately controlled by the direct mixing method.

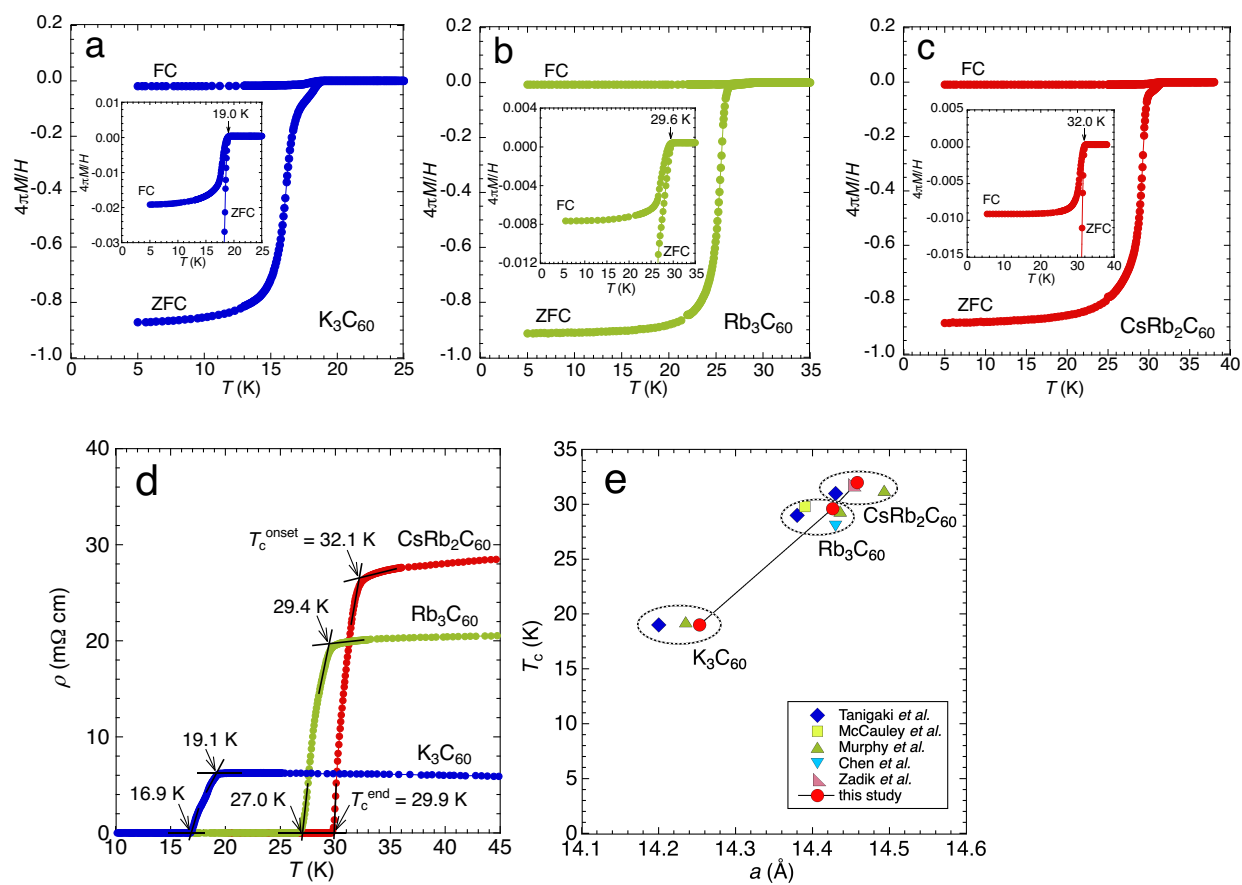

**Figure 4.** Temperature dependence of (a)–(c) $4\pi M/H$ and (d) $\rho$ for $K_3C_{60}$, $Rb_3C_{60}$, and $CsRb_2C_{60}$ synthesized by heat treatment at 500°C for 1 h, respectively. $M$ is corrected using the demagnetization field coefficient based on sample geometry. (e) Relationship between the lattice constants and $T_c$ of $A_3C_{60}$ obtained in this study, along with previously reported data [4,21,22,26].

Table 1. Lattice constants of $A_3C_{60}$ ($A$ = K, Rb, and $Cs_{1/3}Rb_{2/3}$) and $A_6C_{60}$ ($A$ = K, Rb, and Cs) synthesized in this study, along with $T_c^{onset}$, $\Delta T_c$ (= $T_c^{onset}$ - $T_c^{end}$) for $A_3C_{60}$, as determined from resistivity measurements.

| Materials | $a$ (Å) | $T_c^{onset}$ (K) | $\Delta T_c$ (K) |
|---|---|---|---|
| $K_3C_{60}$ | 14.253(1) | 19.1 | 2.2 |
| $Rb_3C_{60}$ | 14.426(1) | 29.4 | 2.4 |
| $CsRb_2C_{60}$ | 14.458(1) | 32.1 | 2.2 |
| $K_6C_{60}$ | 11.380(1) | – | – |
| $Rb_6C_{60}$ | 11.546(1) | – | – |
| $Cs_6C_{60}$ | 11.793(1) | – | – |

## 4. Conclusion

In this study, we developed a practical and efficient method for synthesizing $A_3C_{60}$ using precursors in which $C_{60}$ is pre-intercalated with $A$ through mixing. Powder XRD analysis demonstrated that $A_3C_{60}$ rapidly crystallizes upon heating the precursor above 200–300°C for just 1 h. This short formation period eliminates the need for sealed containers to confine $A$ vapor, simplifying the production process and making it highly suitable for large-scale production. Despite its simplicity, the resulting $A_3C_{60}$ samples exhibited sharp superconducting transitions with large shielding volume fractions. The relationship between lattice constant and $T_c$ was consistent with previous reports. The sinterability of $A_3C_{60}$ achieved through this method is crucial for applications such as superconducting wires and bulk magnets. The simplicity and reliability of this direct mixing method are expected to accelerate research and development for practical applications utilizing $A_3C_{60}$ superconductors.

**Acknowledgements**

This work was supported by the Japan Society for the Promotion of Science (JSPS) KAKENHI Grant Number 22K04193.